\documentclass[aps,prd,twocolumn,groupedaddress,nofootinbib]{revtex4-1}
\pdfoutput=1 

\usepackage[utf8]{inputenc}
\usepackage{mathtools,slashed}
\usepackage[caption=false]{subfig}
\usepackage{dcolumn}
\usepackage{graphicx}
\usepackage{multirow,tabularx,booktabs}
\usepackage{bm}
\usepackage{comment}
\usepackage{setspace}
\usepackage[dvipsnames]{xcolor}
\usepackage[normalem]{ulem} 
\usepackage{enumerate}
\usepackage{siunitx}

\newcommand{\TeV}{{\, {\rm TeV}}}

\newcommand{\vEW}{v_\textsc{ew}}

\newcommand{\LH}{\Lambda_{\mathcal H}}

\newcommand{\Lf}{\Lambda_f}
\newcommand{\fNL}{f_\mathrm{NL}}

\newcommand{\Pz}{\mathcal P_\zeta}
\newcommand{\oa}{\mathsf a}
\newcommand{\ob}{\mathsf b}

\newcommand{\mut}{\widetilde{\mu}}
\newcommand{\lt}{\widetilde{\lambda}}

\definecolor{mypurple}{RGB}{164,64,214}

\usepackage{hyperref}

\begin{document}

\title{Minimal signatures of the Standard Model in non-Gaussianities}
\author{Anson Hook,}\email{hook@umd.edu}\affiliation{Maryland Center for Fundamental Physics, University of Maryland, College Park, MD 20742}
\author{Junwu Huang} 
\email{jhuang@perimeterinstitute.ca}
\affiliation{Perimeter Institute for Theoretical Physics, Waterloo, Ontario N2L 2Y5, Canada}
\author{Davide Racco}\email{dracco@perimeterinstitute.ca}
\affiliation{Perimeter Institute for Theoretical Physics, Waterloo, Ontario N2L 2Y5, Canada}

\begin{abstract}
\noindent
We show that the leading coupling between a shift symmetric inflaton and the Standard Model fermions leads to an induced electroweak symmetry breaking due to particle production during inflation, and as a result, a unique oscillating feature in non-Gaussianities. In this one parameter model, the enhanced production of Standard Model fermions dynamically generates a new electroweak symmetry breaking minimum, where the Higgs field classically rolls into. The production of fermions stops when the Higgs expectation value and hence the fermion masses become too large, suppressing fermion production.  
The balance between the above-mentioned effects gives the Standard Model fermions masses that are uniquely determined by their couplings to the inflaton. In particular, the heaviest Standard Model fermion, the top quark, can produce a distinct cosmological collider physics signature characterised by a one-to-one relation between amplitude and frequency of the oscillating signal, which is observable at future 21-cm surveys.
\end{abstract}

\maketitle

\section{Introduction and summary}

Cosmological collider physics \cite{Chen:2009zp,Baumann:2011nk,Arkani-Hamed:2015bza,Lee:2016vti,Chen:2016hrz,Kumar:2017ecc,Alexander:2019vtb,Lu:2019tjj} provides an opportunity to search for new heavy particles that are not accessible at particle colliders. 
These heavy particles, produced through their interactions with the inflaton field, can accumulate to large enough densities and affect the bispectrum of density perturbations, leaving observable signatures in the cosmic microwave background and large scale structure of the universe \cite{Munoz:2015eqa,Meerburg:2016zdz}.

Given the exciting potential of this approach, it is worth asking ``What are the minimal signatures of the Standard Model (SM) in the context of cosmological collider physics?''.  
Of course the absolutely minimal signature is nothing if the SM does not couple directly to the inflaton. 
In this article, we explore the consequences of adding a single coupling to the SM fermions, the dimension five shift symmetric coupling between the inflaton and the SM fermions. 
Somewhat surprisingly, this single coupling by itself leads to an interplay between the dynamics of the SM fermions and the Higgs: The SM fermion production induces electroweak symmetry breaking during inflation, whereas a large expectation value of the Higgs field increases the masses of the SM fermions and suppresses their production.
This interplay results in a very predictive scenario where the strength of the oscillating signature in non-Gaussianities is directly tied to the period of oscillation.

We assume that the fermions in the Standard Model couple to the inflaton through the lowest dimensional operator respecting the shift symmetry of the inflaton~\cite{Chen:2018xck,Hook:2019long}
\begin{equation}
\label{eq: coupling}
\mathcal{L}\supset 
  \frac{\partial_\mu \phi}{\Lf} \Big(F^{\dagger} \overline \sigma^\mu F + f^{c \, \dagger} \overline\sigma^\mu f^c \Big)
  + y_f \mathcal H F f^c,
\end{equation}
where $ \phi$ is the inflaton, $ \mathcal{H}=\left(0,\frac{v+h}{\sqrt{2}}\right)$ is the Higgs doublet, and $F = Q,\, L$ and $f^c = u^c,\, d^c,\, e^c$ are left- and right-handed fermions in the SM in two-component notation\footnote{
We neglect the typically smaller dimension five anomalous couplings to the gauge bosons.}.
When $\langle \dot\phi \rangle \neq 0$, this coupling leads to the production of fermions during inflation whose effective number density is $n_f \sim m_f \lambda_f^2 \exp[-\pi m_f^2/\lambda_f H]$, where $\lambda_f = \frac{\dot \phi}{\Lambda_f}$, $m_f = y_f v/\sqrt 2$ and $v/\sqrt 2= \left\langle \mathcal{H} \right\rangle$ is the vacuum expectation value (vev) of the Higgs field.

It is well known that a high density of particles can change the properties of a scalar potential. 
Thermal effects are well known to favor symmetry restoration of the Higgs for temperatures above the electroweak phase transition~\cite{Linde:2005ht}.  
On the other hand, chemical potentials favor symmetry breaking and in the context of the Higgs potential could prevent symmetry restoration even at high temperatures~\cite{Benson:1991nj}.  
The coupling in Eq.~\eqref{eq: coupling} is very similar to a chemical potential.  
In fact, if the plus sign were a minus sign, it would be the familiar chemical potential for fermion number.  
As such, it is unsurprising to find that the effect of this coupling is to generate a correction to the Higgs potential that favors symmetry breaking of the form 
\begin{equation}
\delta V_h = - \frac{y_f^2}{2\pi^2} \lambda_f^2 h^2 \exp \left[- \frac{\pi y_f^2 h^2}{2\lambda_f H}\right].
\end{equation}
\begin{figure}[h!] \centering
$\hbox{ \convertMPtoPDF{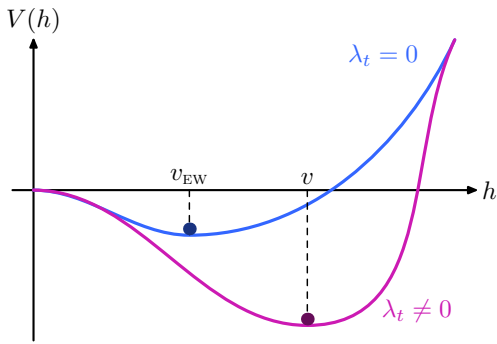}{.8}{.8} }$
\caption{Higgs potential at zero temperature and $\lambda_f\sim 0$ (blue line) and in presence of a top quark condensate induced by $\lambda_t\>H>\vEW$ (purple line).
The Higgs field sits in the dynamically generated minimum $v$ during inflation.
}
\label{fig: Higgs potential}
\end{figure}
For $\lambda_f > H \gg \vEW$ ($\vEW$ being the electroweak vacuum), this contribution to the potential induces spontaneous breaking of the electroweak symmetry, as shown in Fig.~\ref{fig: Higgs potential}. 
However, as the Higgs vev increases, the fermion masses increase and their particle production is exponentially suppressed as $m_f^2 \gtrsim \lambda_f H$.
Therefore, quite insensitively to any other term of the Higgs potential, e.g.\ the value of the quartic $\lambda_h$, the Higgs gets a vev during inflation
\begin{equation}
v \sim \frac{1}{y_f} \sqrt{\lambda_f H} \,.
\end{equation}
In the SM, due to the large hierarchy between the top quark and the lighter leptons and quarks, this effect is determined entirely by the top quark. 
Incidentally, such a scenario can only occur for fermions with an $\mathcal{O}(1)$ Yukawa coupling, since $\lt_f = \lambda_f/H$ cannot be arbitrarily large. 
Thus, in what follows, we focus on the coupling of the top quark with the inflaton.

\begin{figure}[h!] \centering 
\includegraphics[width=0.67\columnwidth]{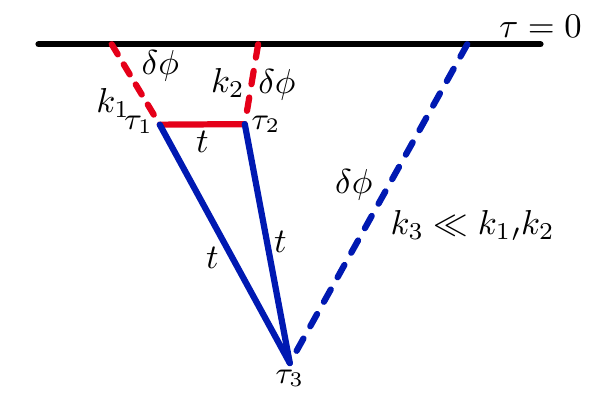}
\caption{Contribution to the inflaton bispectrum from a loop of SM fermions.
Two fermions $f$ are produced at a time $\tau_3$ by the interaction with a soft inflaton leg $\delta\phi$, and annihilate later at $\tau_1 \sim \tau_2$ into two hard inflaton legs with $k_1,k_2\gg k_3$.
The non-analytic contribution to the bispectrum is due to the time propagation of the fermions from $\tau_3$ to $\tau_1$, $\tau_2$.}
\label{fig:dinf}
\end{figure}

The observational signature associated to this coupling is the generation of a large non-Gaussian oscillating pattern in the squeezed limit.  
The Feynman diagram to be calculated is shown in Fig.~\ref{fig:dinf} \cite{Chen:2018xck,Hook:2019long}. 
\newline
The interesting feature in this case is that the dynamics of the Higgs potential ensures that the mass of the fermion $f$ that produces the largest observable non-Gaussianity is solely determined by the inflaton coupling $\lambda_f$ to that fermion. 
As a result, the non-Gaussian signal in the squeezed limit simplifies into
\begin{equation}
\fNL^{(\text{clock})}  \approx  \frac{4\sqrt{2} N_c \Pz}{3 e} \lt_f^{13/2} ,
\label{eq:fNL higgslift}
\end{equation}
depending only on the size of the inflaton fermion coupling in Hubble units $\lt_f=\frac{\lambda_f}{H}=\frac{\dot \phi}{\Lf H}$.  
This signal oscillates in $\ln(k_3/k_1)$ with a frequency $\sim \lt_f$ in the squeezed limit. This relation between the amplitude ($\fNL^{(\text{clock})} $) and the frequency of the oscillating signal offers a simple cross check of this mechanism.
Such a signature offers a direct probe of the {\it induced} electroweak symmetry breaking during inflation, and could help us to shed light on the inflationary sector.

The rest of paper is organized as follows. 
In Sec.~\ref{sec:production}, we discuss the fermion particle production during inflation and calculate the subsequent effect on the Higgs potential. 
In Sec.~\ref{sec:equilibrium}, we study the dynamical equilibrium reached during inflation and how the fermion mass is set by its coupling to the inflaton. 
In Sec.~\ref{sec:signature}, we discuss the signature in the bispectrum of density perturbations ultimately sourced by the dynamically generated Higgs vev, and the implications for the detection of this signal.

\section{Particle production and the Higgs potential}\label{sec:production}

\begin{figure}[h!] \centering \vspace{-1.em}
\includegraphics[width=\columnwidth]{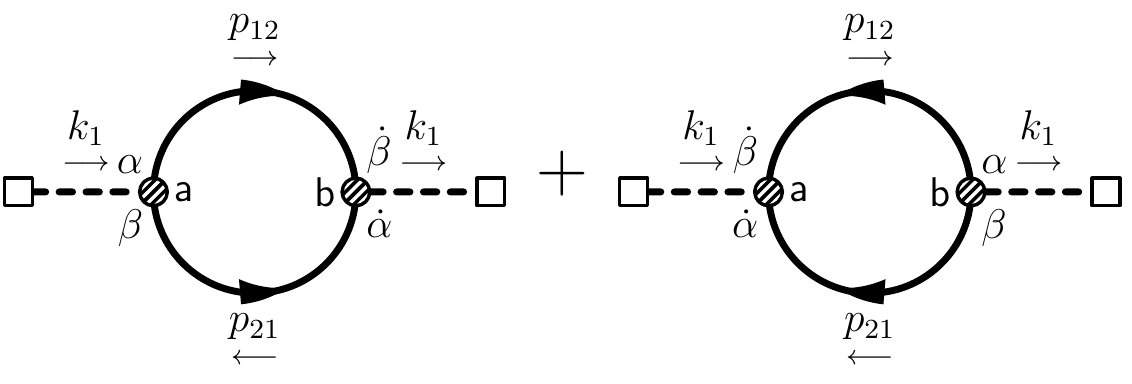}
\vspace{-1.8em}
\caption{Feynman diagrams for the contribution of the top fermion condensate to the Higgs potential. For notations, see the main text and references~\cite{Hook:2019long,Chen:2016nrs,Chen:2017ryl}.}
\label{fig: Higgs loop}
\end{figure}

In this section, we will discuss particle production and how the fermion density affects the Higgs potential. 
To calculate the correction to the Higgs mass term, we calculate the standard diagram shown in Fig.~\ref{fig: Higgs loop} that corrects the energy density of the state.  

In a companion paper~\cite{Hook:2019long} (see in particular Sec.~3 and Appendix~A), we show in detail how to estimate and calculate these diagrams. 
For a single fermion flavor and color, the diagrams in Fig.~\ref{fig: Higgs loop} contribute as
\begin{widetext}
\begin{equation}
\begin{aligned}
 N_{\oa\ob} = & - y_f^2\epsilon^{\alpha \beta} \epsilon^{\dot \alpha \dot \beta}
 \oa\ob \iint_{-\infty}^0 \frac{ \mathrm{d} \tau_1}{H\tau_1} \frac{\mathrm{d} \tau_2}{H\tau_2} G_{\oa}(\vec k_1,\tau_1) G_{\ob}(-\vec k_1,\tau_2) \\
& \times \int  \frac{\mathrm d^3q}{(2\pi)^3} 
\left(
  D_{\oa \ob \alpha \dot \beta}(\vec p_{12},\tau_1,\tau_2)
  D_{\oa \ob \beta \dot \alpha}(-\vec p_{21},\tau_1,\tau_2)+
  D_{\ob \oa \alpha \dot \beta}(-\vec p_{12},\tau_2,\tau_1)
  D_{\ob \oa \beta \dot \alpha}(\vec p_{21},\tau_2 ,\tau_1)\right),
\end{aligned}
\label{eq:diagram}
\end{equation}
where $\tau$ denotes conformal time, $\vec p_{12} = \vec k_1 + \vec p_{21} = \vec q$ and $|\vec q |\gg |\vec  k_{1}|$ is the internal momentum. 
The indices $\alpha,\,\beta,\, \dot \alpha,\,\dot \beta \in \{1,2\}$ are spinor indices. 
The antisymmetric $\epsilon$ tensors are defined by $\epsilon^{12} = -\epsilon_{12} = 1$. 
The in-in indices $\oa,\, \ob$ take values in $\{+1,-1\}$, denoted respectively by $\{\oplus,\ominus\}$ to distinguish them from spinor indices. 
The functions $G_{\oa}(\vec k,\tau) =\sqrt{H^2/2k^3}(1-i\oa k\tau)e^{i\oa k\tau}$ and $D_{\oa \ob \alpha  \dot \beta}(\vec k,\tau_1,\tau_2)$ are the propagator of the Higgs and the fermion fields respectively, and can be found in Sec.~A.2 of \cite{Hook:2019long}.
By the same logic of the Electroweak Hierarchy problem, the momentum integral in Eq.~\eqref{eq:diagram} is quadratically divergent. 
The leading quadratic divergence is absorbed into the definition of the physical Higgs mass that we observe today when $\lambda_f=0$ (this operation automatically removes some of the subleading corrections in the $\lt_f$ expansion). 
In the massless limit, the integral in Eq.~\eqref{eq:diagram} can be simplified (by exploiting properties of the Whittaker functions appearing in the fermion mode functions) to be
\begin{equation}
N_{\oa\oa} = y_f^2 \int \frac{\mathrm d^3q}{(2\pi)^3} \iint \frac{ \mathrm{d} \tau_1}{H\tau_1} \frac{\mathrm{d} \tau_2}{H\tau_2} 2\theta(\tau_2-\tau_1)e^{- i \oa (p_{12} +p_{21})(\tau_2 -\tau_1)} \left(\left(\frac{\tau_1}{\tau_2}\right)^{2 i\lt_f}+\left(\frac{\tau_2}{\tau_1}\right)^{2 i\lt_f}\right) G_{\oa}(\vec k_1,\tau_1) G_{\oa}(-\vec k_1,\tau_2) + \cdots
\label{eq:masslessdiagram}
\end{equation}
\end{widetext}
where we only consider the terms where $\oa=\ob$.  The terms with $\oa \ne \ob$ are not divergent, not unrelated to the fact that the counterterm $\delta m^2\,h^2/2$ for the Higgs mass can only cancel the divergences in $\delta m^2_{\oplus\oplus}$ ($N_{\oplus\oplus}$) and $\delta m^2_{\ominus\ominus}$ ($N_{\ominus\ominus}$).
Moreover, these two counterterms are related by a minus sign (due to complex conjugation).
As a result, we can find the correction to the Higgs potential from the diagrams in Fig.~\ref{fig: Higgs loop} by taking the difference between the two diagrams, removing in this way the corrections that are linearly divergent~\cite{Chen:2016nrs}
\begin{equation}
\frac{\delta m^2}{3 k^3} =\frac{1}{3 k^3} \frac{\delta m^2_{\oplus\oplus}-\delta m^2_{\ominus\ominus}}{2} = -\frac{y_f^2 \lambda_f^2}{3 \pi^2 k^3}.
\end{equation} 
This gives the correction from a single massless fermion flavor/color to the Higgs potential 
\begin{equation}
\delta V_h = - \frac{y_f^2}{2\pi^2} \lambda_f^2 h^2 
\end{equation}
at leading order in the large $\lt_f$ expansion%
\footnote{In the massless limit, the potential can be computed to next-to-leading-order in large $\lt_f$ expansion to be $\delta V_h = - \frac{y_f^2}{2\pi^2} \left(\lambda_f^2 -\frac{H^2}{2}\right)h^2$.} and to leading order in the Higgs field. 

In the large $m_f$ limit, we obtain an exponential suppression with exponent $\sim -y_f^2 h^2/\lambda_f H$. 
We cannot calculate the coefficient of the exponent.  As an approximation, we match this exponential suppression for the correction to the Higgs potential to the exponential suppression of the fermion density%
\footnote{We acknowledge the potential existence of subleading terms in the small $m_f/H$ limit due to our inability to resum and reproduce the exact exponential suppression in the fermion density.}.
 
There is not a clean analytic formula interpolating between the small and large mass regions and a full result would need to be obtained numerically. 
For simplicity, we interpolate between the two expressions using the following potential in the $SU(2)_L$ symmetric form:
\begin{equation}
V_h = -\mu_h^2 |\mathcal{H}|^2 +\lambda_h |\mathcal{H}|^4- \frac{N_c y_f^2}{\pi^2} \lambda_f^2 |\mathcal{H}|^2 \exp \left[-\frac{ \pi y_f^2 |\mathcal{H}|^2}{\lambda_f H}\right],
\label{eq: V(h)}
\end{equation}
up to corrections that are independent of $\lambda_f$, where $\mu_h$ and $\lambda_h$ are the coefficients of the quadratic and quartic terms of the Higgs potential, and $m_h=\sqrt 2 \mu_h$ is the physical mass of the Higgs boson.

\section{Non-Gaussian Signature}
\label{sec:equilibrium}
The induced Higgs potential in Eq.~\eqref{eq: V(h)} has a new minimum at
\begin{equation}
v = \frac{1}{y_f} \sqrt{\frac{2}{\pi} \lambda_f H}\left(1 - \frac{e \pi \lambda_h/y_f^4}{ N_C \lt_f } + \mathcal{O}(\lambda_h^2)\right).
\label{eq: v Higgs}
\end{equation}
for $v\gg \vEW$. In the SM, the top quark Yukawa is $y_t = \sqrt{2} m_t/v_{\rm EW} \approx 1$ at the weak scale and runs to about $0.6$ at very high energies, while the Higgs quartic is $\lambda_h = m_h^2/ 2 v_{\rm EW}^2 \approx 0.13$ at the weak scale and runs to $\lambda_h \lesssim 0.01$ at very high energies. This ensures that the small quartic expansion is a very good approximation for $\lambda_f/H \gg 1$.

\begin{figure}[h!] \centering 
\includegraphics[width=\columnwidth]{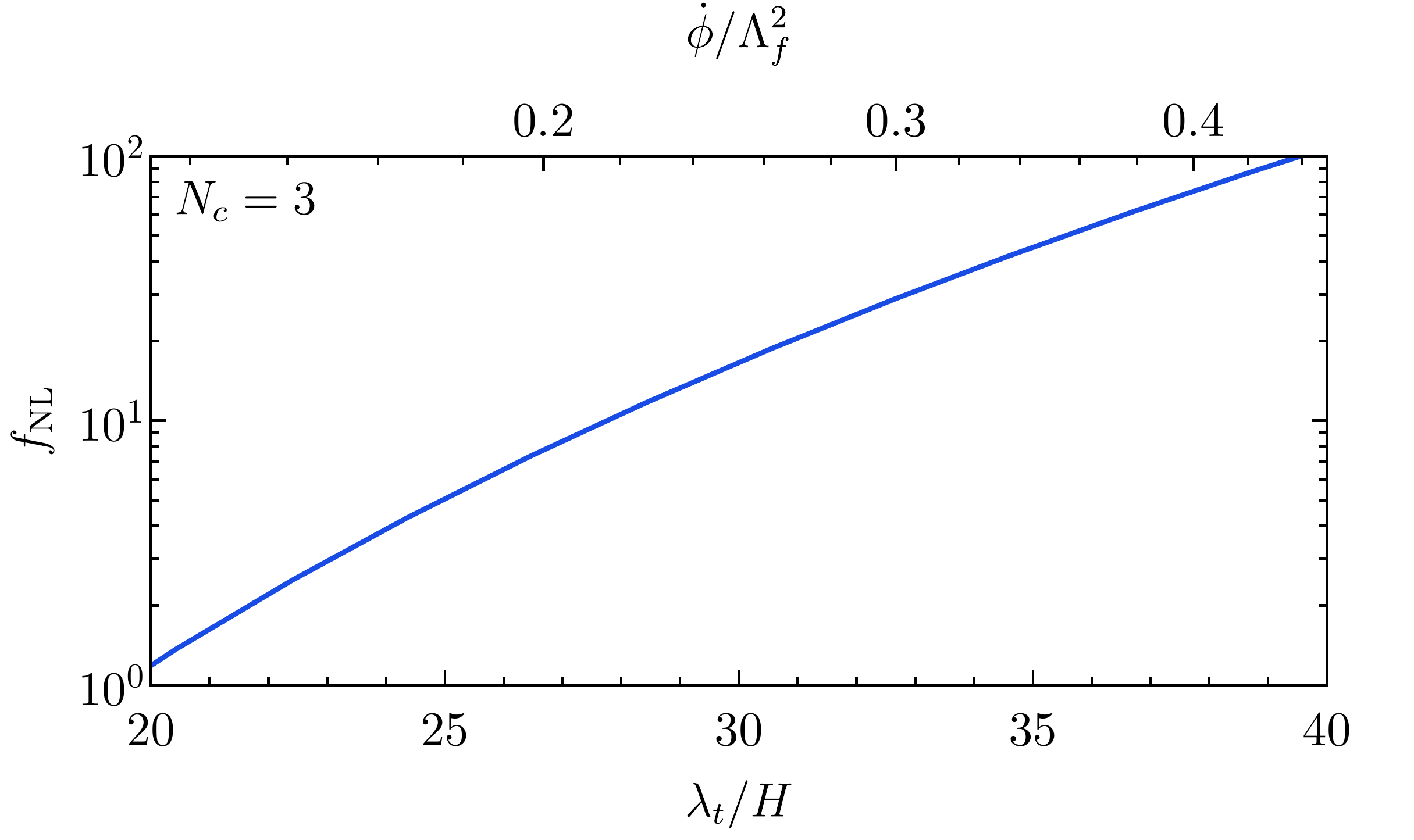}
\caption{Size of $\fNL$ generated by the top quark condensate in the scenario described in this paper (Eq.~\ref{eq: fNL top}), as a function of $\lambda_t/H=c_t \dot\phi/\Lf H$ (or $\dot\phi/\Lf^2$, assuming $c_t=1$).}
\label{fig: fNL}
\end{figure}

In \cite{Hook:2019long}, we discuss in detail how to compute the non analytic signal induced in the bispectrum by fermions interacting with the inflaton. 
The result in the squeezed limit and at leading order in the large $\lt_f$ expansion is
\begin{equation}
S(k_1,k_2, k_3) \overset{k_3 \ll k_1 \sim k_2} \simeq f_{\rm NL}^{({\rm clock})} \left( \frac{k_3}{k_1}\right)^{2-2 i \lt_f},
\end{equation}
where $m_f =\sqrt{\lambda_f H/\pi} \ll \lambda_f$, and
\begin{multline}
f_{\rm NL}^\text{(clock)} \approx \frac{N_c}{6\pi} \Pz^{-1/2} \left(\frac{m_f}{\Lf}\right)^3 \lt_f^2 \\ 
 \cdot \frac{e^{\pi \lt_f} \mut_f \Gamma (-i \mut_f )^2 \Gamma (2 i \mut_f )^3}{2 \pi  \Gamma (i(\lt_f +\mut_f ))^3 \Gamma (i (\mut_f -\lt_f )+1)} \,,
\end{multline}
where $\mu_f=\sqrt{\lambda_f^2+m_f^2}$ and $\mut_f= \mu_f/H$. In the scenario discussed in this paper, the contribution to the Higgs potential from the top quark condensate {\it dynamically adjusts} the Higgs vev so that $m_f =\sqrt{\lambda_f H/\pi}$, and the non-analytic contribution to the bispectrum is just a function of $\lt_f$:
\begin{equation}
f_{\rm NL}^\text{(clock)} \approx \frac{4\sqrt{2} N_c \Pz}{3 e} \lt_f^{13/2}.
\label{eq: fNL top}
\end{equation}
We show the contribution to $\fNL$ from the top quark in our scenario as a function of $\lambda_t/H$ in Fig.~\ref{fig: fNL}.
\vspace{-1.em}
\newline

\section{Remarks}
\label{sec:signature}

In this article, we described the signatures of shift symmetric couplings between the inflaton and the SM, in particular the one to the top quark.  
This coupling dynamically sets the masses of the SM fermions and simultaneously determines the strength of the non-Gaussianity giving rise to the unique feature that the frequency of the oscillating pattern in $\ln(k_3/k_1)$ fixes the amplitude of the non-Gaussian signal.
This provides a way to distinguish the scenario that we describe in this paper from other models generating this feature.

Surprisingly, high energy colliders probing the flavor structure of the SM~\cite{Dolan:2014ska} provide an alternative route for testing our scenario. 
For a typical slow-roll inflation model~\cite{Guth:1980zm,Linde:1981mu,Lyth:1998xn,Baumann:2009ds}, the inflaton mass $m_\text{inf}$ is%
\footnote{We assume for simplicity a scenario where the inflaton mass today does not differ too much from the inflaton mass $m_\text{inf}$ during inflation.} 
\begin{equation}
m_{\rm inf} \ll H_{\rm inf} \lesssim 10^{-2} \Lf \,,
\end{equation} 
where the last inequality comes from the requirement that $\lt_f=\frac{\Pz^{-1/2}}{2\pi} \frac{H}{\Lf}=\frac{\Pz^{-1/4}}{\sqrt{2\pi}} \left(\frac{\dot\phi}{\Lf^2}\right)^{1/2}$ satisfies $\dot\phi\lesssim \Lf^2$ so that we can rely on the EFT expansion in powers of $\Lf$. 
Experimental constraints from BaBar and exotic meson decays imply a bound $\Lf \gtrsim 10\TeV$ for scalars that couple with ``Yukawa-like couplings'' to the three generations of SM fermions (see Secs.~4.1 and 4.2 in~\cite{Dolan:2014ska}). 
A detailed study of the flavor signatures of this scenario is beyond the scope of this paper.

A potentially observable signal in future measurements of the  bispectrum (say $\fNL\gtrsim 10$) requires a large $\lt_f$, which together with $\Lf\gtrsim 10\TeV$ translates to a lower bound on the Higgs vev during inflation: 
\begin{equation}
v \approx \frac{1}{y_t} \sqrt{\frac{2 \lambda_f H}{\pi}} \approx \frac{\sqrt{8\pi}}{y_t} \lt_f^{3/2} \Pz^{1/2} \Lf \gtrsim 300 \,{\rm GeV} .
\end{equation}
Similarly to the correction to the Higgs vev during inflation from $\lambda_h$, there is also a correction from $v_{\rm EW}\equiv \sqrt{m_h^2/2\lambda_h}$ when $\lambda_f \sim v_{\rm EW}$. Such a correction amounts to an expansion in $\frac{\vEW^2}{v^2} \frac{e \pi \lambda_h}{ y_t^4  N_c\lt_f}$, which is also a small correction as long as $v \gtrsim \vEW$.

A potentially significant correction to the scenario we have described can arise if there is a significant coupling between the inflaton and the Higgs boson in the form of
\begin{equation}
\label{eq: last}
c_2 \left(\frac{\partial \phi}{\LH}\right)^2 \mathcal{H}^{\dagger} \mathcal{H}.
\end{equation}
This coupling can be generated from UV dynamics or by integrating out SM fermions that couple directly to the inflaton. 
In order for the scenario described in this paper to follow through, we needed to be able to suppress any large corrections to the Higgs mass.  
During inflation, Eq.~\eqref{eq: last} leads to a Higgs mass term of order $c_2 (\dot \phi/\LH)^2$, potentially comparable to the mass term shown in Eq.~\eqref{eq: v Higgs}. 
This mass term (like the term $-\mu_h^2 |\mathcal H|^2$ in Eq.~\eqref{eq: V(h)}) can lead to corrections to the Higgs vev $v$ during inflation. 
Requiring that Eq.~\eqref{eq: last} is a negligible correction to $v$ implies $c_2 \lesssim \frac{ N_c y_t^2}{\pi^2e} \approx 0.1$ if $\Lf = \LH$. 
If $c_2$ is large and positive, the minima in the Higgs potential would disappear (together with the signal), while if $c_2$ is large and negative, then we go back to the cases studied in~\cite{Kumar:2017ecc,Hook:2019long}, where the Higgs vev is not dynamically adjusted by the mechanism discussed here.

The signature that we discuss can also arise whenever there is a fermion whose mass is generated from an $\mathcal{O}(1)$ Yukawa coupling.  For example, if the dark sector contains dark fermions whose mass is generated (also partially) from Yukawa couplings to a dark Higgs boson, then the leading order shift symmetric couplings between the dark fermions and the inflaton would result in a signal similar to the one we discuss (with a different $N_c$).

Soon, the limits on non-Gaussanities and the scalar-to-tensor ratio will improve dramatically~\cite{Abazajian:2016yjj,Meerburg:2016zdz,Munoz:2015eqa,Shandera:2010ei}.  
It is thus extremely interesting to ask what sort of signals might be discovered in the upcoming years.  
In this article, we described the most minimal way in which the SM itself may reappear in the sky.  In the near future, a combination of flavor measurements at particle colliders and bispectrum measurements at the cosmological collider may combine to uncover deep mysteries of the universe.

\begin{acknowledgments}
We would like to acknowledge Asimina Arvanitaki for relentlessly inviting us to think about this scenario. 
We also thank Prateek Agrawal, David Curtin, Savas Dimopoulos, Jiji
Fan, Daniel Ega\~na-Ugrinovic, Matthew Johnson, Soubhik Kumar, Gustavo Marques-Tavares, Maxim Pospelov, Antonio Riotto, Raman Sundrum, Zhong-Zhi Xianyu for many useful discussions. 
JH and DR thank the Stanford Institute for Theoretical Physics, and AH and JH thank MIAPP and KITP Santa Barbara, for generous hospitality during the completion of this work.
This research was supported in part by Perimeter Institute for Theoretical Physics. Research at Perimeter Institute is supported in part by the Government of Canada through the Department of Innovation, Science and Economic Development Canada and by the Province of Ontario through the Ministry of Economic Development, Job Creation and Trade.  
AH is supported in part by the NSF under Grant No.\;PHY-1620074 and by the Maryland Center for Fundamental Physics (MCFP). This research was supported in part by the National Science Foundation under Grant No. NSF PHY-1748958.
\end{acknowledgments}

\bibliographystyle{apsrev4-1}
\bibliography{higgslift_Refs}
\end{document}